\RequirePackage{ifpdf} 
\ifpdf
\documentclass[pdftex,12pt,a4paper]{article}  
\else
\documentclass[12pt,a4paper]{article} 
\fi

\parskip=\medskipamount

\usepackage[mathscr]{euscript} 	
\usepackage{bbm} 
\usepackage{amsfonts,amsmath,amssymb}
\ifpdf
\usepackage[pdftex]{graphicx}	
\usepackage[usenames,pdftex]{color}		
\usepackage[pdftex,bookmarks]{hyperref}
\hypersetup{bookmarksopen,bookmarksnumbered,
		pdfnewwindow=true,
		pdfauthor = {Simon J. Tyler},
		pdftitle = {Complex linear superfield as a model for Goldstino},
		pdfsubject = {},
		pdfkeywords = {Supersymmetry, Goldstino},
		pdfcreator = {LaTeX with hyperref package},
		pdfproducer = {pdflatex}
}
\else
\usepackage[dvips]{graphicx}
\usepackage[usenames,dvips]{color}
\usepackage[dvips,bookmarks,colorlinks=false]{hyperref}
\fi
\def\a{\alpha}
\def\b{\beta}
\def\c{\chi}
\def\d{\delta}

\def\f{\phi}
\def\g{\gamma}
\def\G{\Gamma}
\def\h{\eta}

\def\j{\psi}
\def\k{\kappa}
\def\l{\lambda}

\def\q{\theta}
\def\r{\rho}
\def\s{\sigma}

\def\w{\omega}
\def\x{\xi}

\def\F{\Phi}

\def\L{\Lambda}

\def\S{\Sigma} 

\def\X{\Xi}
\newcommand{\eps}{\varepsilon}
\newcommand{\vf}{\varphi}
\newcommand{\da}{{\dot{\alpha}}}
\newcommand{\db}{{\dot{\beta}}}

\newcommand{\ada}{{\alpha\dot\alpha}}

\def\ds1{\ensuremath{\mathbbm{1}}}
\newcommand{\rmd}{{\rm d}}
\newcommand{\rme}{{\rm e}}
\newcommand{\rmi}{{\rm i}}

\newcommand{\cD}{{\cal D}}
\newcommand{\cDb}{\bar\cD}

\newcommand{\cN}{{\cal N}}
\newcommand{\st}{{\tilde\s}}

\newcommand{\expt}[1]{\ensuremath{\big< #1 \big>}}

\newcommand{\pd}{\partial}
 				
\def\tr{{\rm tr}}		
\def\const{{\rm const}}
\def\cc {{\rm c.c.}}

\def\intx{\int\!\!{\rmd}^4x\,}

\newcommand{\be}{\begin{equation}}
\newcommand{\ee}{\end{equation}}
\newcommand{\bea}{\begin{eqnarray}}
\newcommand{\eea}{\end{eqnarray}}
\newcommand{\non}{\nonumber}
\newcommand{\bm}[1]{\mbox{\boldmath$#1$}}
\newcommand{\Db}{{\bar{D}}}

\newcommand{\wb}{{\bar{\w}}}

\begin{document}                        
\begin{titlepage}

\begin{flushright} February, 2011 \end{flushright}

\vspace{5mm}

\begin{center}
{\large \bf  Complex linear superfield as a model for Goldstino}
\end{center}
\begin{center}
{\large  
{Sergei M.\ Kuzenko}
\footnote{kuzenko@cyllene.uwa.edu.au}
and 
{Simon J.\ Tyler}
\footnote{styler@physics.uwa.edu.au}

\vspace{5mm}

\footnotesize{
{\it School of Physics M013, The University of Western Australia\\
35 Stirling Highway, Crawley W.A. 6009, Australia}}  

\vspace{2mm}}
\end{center}

\vspace{5mm}

\pdfbookmark[1]{Abstract}{abstract_bookmark}
\begin{abstract}
\baselineskip=14pt
We propose a Goldstino model formulated in terms of 
a constrained complex linear superfield.
Its comparison to other Goldstino models is given.
Couplings to supersymmetric matter and supergravity are briefly described.
\end{abstract}
\vfill

\end{titlepage}


\section{Introduction}\label{sect:Intro}
Since the pioneering work of Volkov and Akulov \cite{VA,AV} 
in which they proposed the Goldstino action, 
there have appeared alternative formulations in which the Goldstino
is described in terms of constrained $\cN=1$ superfields. 
The most famous constructions are: 
(i) Ro\v{c}ek's model \cite{Rocek1978} realized in terms of
	a constrained chiral superfield;
(ii) the Lindstr\"om-Ro\v{c}ek model \cite{LR} realized in terms of
	a constrained real scalar superfield;
(iii) the Samuel-Wess model \cite{SamuelWess1983} which is formulated using
	a constrained spinor superfield.%
\footnote{According to the general theory of the nonlinear realisation of
${\cal N}=1$ supersymmetry \cite{AV,Ivanov,Uematsu:1981rj}, 
the Akulov-Volkov action is universal in the sense that 
any Goldstino model should be related to the Akulov-Volkov action by a 
nonlinear field redefinition. However, the superfield models 
\cite{Rocek1978,LR,SamuelWess1983} are interesting in their own right, 
in particular in the context of supergravity \cite{LR,SamuelWess1983}.} 
What is missing in this list of Goldstino models is a realization involving 
a complex linear superfield.
The present note is aimed at filling this gap.
Our notation and conventions correspond to \cite{BK}.

\section{Constrained complex linear superfield}\label{sect:Constraints}
A complex linear superfield $\G$ obeys  the only constraint ${\bar D}^2 \G =0$,  
and can be used to provide an off-shell description for the scalar multiplet 
({\it non-minimal scalar multiplet}) \cite{BK,GGRS}.
A modified complex linear superfield, $\S$,
is defined to satisfy the constraint
\begin{align} \label{mCL:constraint}
	-\frac14 \Db^2 \S = f\,, \qquad f = {\rm const}\,.
\end{align}
Here $f$ is a parameter of mass dimension 2 which, 
without loss of generality, can be chosen to be real.
The above constraint naturally occurs if one introduces a dual
formulation for the chiral scalar model
\begin{align} \label{effective-action2}
	S[\F,\bar\F]  = \intx\!\rmd^2\q\rmd^2\bar\q\,\bar\F\F
		+ \Big(  f \intx\rmd^2\q\,\F + {\rm c.c.} \Big)			\,,
\end{align}
with  $\F$ being  chiral. 
The general solution to the  constraint \eqref{mCL:constraint} is
\begin{align} \label{mCL:components}
	\S (\q,\bar\q)= \rme^{\rmi\q\s^a\bar\q\pd_a}
		\left(\f + \q\j + \sqrt2\bar\q\bar\r
		+ \q^2F + \bar\q^2f + \q^\a\bar\q^\da U_\ada + \q^2\bar\q\bar\c
		\right)\ .
\end{align}

The free action for the complex linear superfield is 
\begin{align}
	S[\S,\bar\S] 
		&= - \intx\!\rmd^2\q\rmd^2\bar\q\, \S\bar\S  \non \\
		&= -\intx\left(f^2 + \bar F F - \f\Box\bar\f + \rmi\r\pd\bar\r 
		- \frac12\j {\bm \c} - \frac12\bar\j\bar{\bm \c}
		- \frac12 \bar{\bm U}^a\bm{U}_a
		 \right) \,,
\label{mCL-free-action}
\end{align}
where we have introduced
\begin{align}
	{\bm U}^a = U^a + 2\rmi\pd^a\f \,, \qquad 
	{\bm \c} = \c-\frac\rmi2\pd\bar\j\,.
\end{align}
It is seen from the component expression for  $S[\S,\bar\S]$ that $\f$ and $\r$ 
are physical fields while the rest of the fields are auxiliary.

It turns out that the above action is suitable to describe 
the Goldstino dynamics provided $\S$ is subject to 
the following nonlinear constraints:
\begin{align}
	\S^2 &= 0~, \label{1st constraint} \\
	-\frac{1}{4} \S\Db^2D_\a\S &= f D_\a\S~. \label{2nd constraint}
\end{align}
The constraints can be seen to be compatible.  Using \eqref{mCL:constraint}, 
the second constraint can be rewritten in the form:
\begin{align}
	\rmi\S\pd_\ada\Db^\da\S = -fD_\a\S\,.
\end{align}
Any low-energy action of the form
\begin{align}
	S_\mathrm{eff} = 
	\intx\!\rmd^2\q\rmd^2\bar\q\, K(\bar\S,\S) 
\end{align}
reduces to \eqref{mCL-free-action} if $\S$ is subject to the nilpotent 
condition  \eqref{1st constraint}.

The general solution to the constraint  \eqref{1st constraint}
fixes $\f$ and two of the auxiliary fields
\begin{align} \label{soln-to-1st-constraint}
	f \f   = \frac12\bar\r^2\,, \quad
	f\j_\a = \frac1{\sqrt2} U_\ada\bar\r^\da\,, \quad
	f F    = \frac1{\sqrt2}\bar\c\bar\r + \frac14 U^a U_a \ .
\end{align}
Taking into account the second constraint, eq. \eqref{2nd constraint}, 
fixes all of the components as functions of the Goldstino $\bar\r$
\begin{equation}\label{solns-to-both-constraints}
\begin{gathered}	f \f			= \frac12\bar\r^2\,, \quad
	\sqrt2f^2\j_\a	= -\rmi\bar\r^2(\pd\bar\r)_\a \,, \quad
	f^{3}F 			= \bar\r^2(\pd_a\bar\r\st^{ab}\pd_b\bar\r)\,, \\
	f U_\ada		= 2\rmi(\s^a\bar\r)_\a\pd_a\bar\r_\db\,,  \quad
	f^2\bar\c_\da 	= \sqrt2\big((\bar\r\st^a\s^b\pd_b\bar\r)\pd_a\bar\r_\db
						- \frac12 (\Box\bar\r^2)\bar\r_\db \big)\ .
\end{gathered}\end{equation}
Note that the simplicity of these solutions follows from the fact that 
the two constraints depend only on $\S$ and not $\bar\S$.

The Goldstino action that follows from
\eqref{mCL-free-action} and \eqref{solns-to-both-constraints} is
\begin{equation}\label{eqn:chAV_Action}\begin{aligned} 
	S[\r,\bar\r] 
		&=-\frac12\intx\Big(\k^{-2}+\expt{\w+\wb}
			+\k^2\big(\pd^a\r^2\pd_a\bar\r^2+4\expt{\w}\expt{\wb}\big) \\
		&\qquad	+\k^4\big(\expt{\w}\big(2\pd^a\r^2\pd_a\bar\r^2
				+4\expt{\w\wb}+4\expt{\wb}^2-2\expt{\wb^2}
				-\bar\r^2\Box\r^2\big) + \cc\big)	\\
		&\qquad	+\k^6\big(\r^2\bar\r^2\Box\r^2\Box\bar\r^2
				-8\expt{\w}^2\expt{\wb^2}-8\expt{\w^2}\expt{\wb}^2\big)
		\Big)\,,
\end{aligned}\end{equation}
where we have introduced, to ease the comparison with the standard literature on
nonlinearly realized supersymmetry,  
 the coupling constant $\k$ defined by $2\k^2=f^{-2}$.
We have also introduced the notation
\begin{align}
	\w_a{}^b=\rmi\r\s^b\pd_a\bar\r\,,\qquad
	\wb_a{}^b=\rmi\bar\r\st^b\pd_a\r\,,
\end{align}
as well as denoted by $\expt{M}$ the matrix trace of any matrix, $M=(M_a{}^b)$,
with Lorentz indices.
The above action proves to be  the same as 
the component action described by Samuel and Wess \cite{SamuelWess1983} 
(see, e.g.,\ eq.\ (41) of \cite{paper}).
The reason for this will be explained shortly.

It is instructive to compare the constraints \eqref{1st constraint} 
and \eqref{2nd constraint} with those corresponding to Ro\v{c}ek's 
Goldstino action \cite{Rocek1978}. The latter model is described by 
a chiral scalar $\F$, 
\begin{align}
	{\bar D}_\da \F=0\,,
\end{align}
constrained as follows:
\begin{align} 
\label{Roceks_Constraint1}
	\F^2 &= 0 \,,   \\
\label{Roceks_Constraint2}
	-\frac14 \F\Db^2\bar\F &= f\F \,, \qquad f={\rm const}\,, 
\end{align}
where the parameter 
can also be  chosen real.
The constraint \eqref{Roceks_Constraint2} mixes $\F$ and its conjugate $\bar\F$, 
while \eqref{2nd constraint} involves $\S$ only.
The complete solution to the constraints 
(\ref{Roceks_Constraint1}) and (\ref{Roceks_Constraint2})
can be found  in \cite{Rocek1978,paper}.

Naturally associated with $\S$ and $\bar \S$ are 
the spinor superfield $\X_\a$
and its conjugate ${\bar \X}_\da$ defined by 
\begin{align} \label{SW_as_Derivative}
\X_\a = \frac{1}{\sqrt2}D_\a \bar \S\,, \qquad
	\bar\X_\da = \frac{1}{\sqrt2}\Db_\da\S\ .
\end{align}
Making use of the constraints \eqref{mCL:constraint}, \eqref{1st constraint} 
and \eqref{2nd constraint}, we can readily uncover 
those constraints which are obeyed by the above spinor superfields. 
They are
\begin{align}
\label{SW_Constraint1}
	{\bar D}_\da {\bar \X}_\db 
		&= \k^{-1} \eps_{\da \db}\,, \\
\label{SW_Constraint2}
	D_\a {\bar \X}_\da
		&= 2 \rmi \k {\bar \X}^\db \pd_{\a \db} {\bar \X}_\da  \ ,
\end{align}
where, as above,  $2\k^2=f^{-2}$.
These are exactly the constraints given in \cite{SamuelWess1983},
so we recognise $\X_\a$ as the Samuel-Wess superfield.
This connection is discussed in more detail in the next sections.
It appears that the Goldstino realization in terms of $\S$ and $\bar \S$ 
is somewhat more fundamental than the one described by eqs.\ 
(\ref{SW_Constraint1}) and (\ref{SW_Constraint2}).

\section{Comparison to other Goldstino models}\label{sect:comparisons}
The two most basic Goldstino models start with the nonlinear 
Akulov-Volkov (AV) supersymmetry \cite{AV}
\begin{align} \label{AV:SUSY}
	\d_\h\l_\a &= \frac1\k\h_\a 
			-\rmi\k \big(\l\s^b\bar\h-\h\s^b\bar\l\big)\pd_b\l_\a \,,
\end{align}
and the chiral nonlinear AV supersymmetry
\begin{align} \label{chAV_SUSY}
	\d_\h\x_\a = 
		\frac1\k\h_\a - 2\rmi\k(\x\s^a\bar\h)\pd_a\x_\a \ .
\end{align}
The latter supersymmetry first appeared in \cite{Zumino:ChiralNLSusy} 
before being discussed in \cite{Ivanov} and \cite{Rocek1978}. 
It was then central to the approach of Samuel and Wess
\cite{SamuelWess1983} that we discuss below.

As discussed in \cite{Luo:2009ib}, the AV supersymmetry is naturally associated
with a real scalar superfield 
(also known as ``vector superfield'' in the early supersymmetry literature), 
while the chiral AV supersymmetry
is associated with a chiral scalar.
Constraints that eliminate all fields but the Goldstino have previously been 
given for both of these types of superfields.
The first \cite{Rocek1978} was for the chiral scalar, $\F$, where Ro\v{c}ek
introduced the constraints 
\eqref{Roceks_Constraint1} and \eqref{Roceks_Constraint2}.
The relevant constraints for the real scalar, 
\begin{align} \label{RL_Constraints}
	V^2=0\,,  \quad \quad VD^\a\Db^2D_\a V=16fV \,,
\end{align}
were given by Lindstr\"om and Ro\v{c}ek \cite{LR}.
The first constraint in both of these sets is a nilpotency constraint, 
while the second is such that the free action is equivalent to 
a pure $F$- or $D$-term respectively.
This latter property is not one possessed by the second constraint 
\eqref{2nd constraint} for the complex linear superfield.

The constraints for both the chiral and real scalar superfields were solved in
\cite{SamuelWess1983} in terms of the spinor Goldstino superfield 
\begin{align} \label{chiral-Goldstino-SF}
	\X_\a(x,\q,\bar\q) = \rme^{\d_\q}\x_\a (x)\  .
\end{align}
The actions of the supercovariant derivatives $D_\a$ and $\Db_\da$ on $\X_\a$
follow from the supersymmetry transformation \eqref{chAV_SUSY}
and are exactly the constraints presented in
\eqref{SW_Constraint1} and (\ref{SW_Constraint2}).
The solutions for the constrained superfields that were given in 
\cite{SamuelWess1983} are
\begin{align} \label{SW-solns}
	2f\F = -\frac{\k^2}{4}\Db^2\big(\X^2\bar{\X}^2\big)\,,\qquad
	2fV = {\k^2}\X^2\bar{\X}^2 \ .
\end{align}
From these solutions, it is straightforward to check that $fV=\F\bar\F$.

It is interesting to note that exactly the same solutions work, 
\begin{align} \label{SW-solns2}
	2f\F = -\frac{\k^2}{4}\Db^2\big(\L^2\bar{\L}^2\big)\,,\qquad
	2fV = {\k^2}\L^2\bar{\L}^2 \ ,
\end{align}
if we replace $\X$ with the spinor
Goldstino superfield that follows from the normal AV supersymmetry
(see, e.g., \cite{WB})
\begin{align} \label{AV_Superfield}
	{\L}_\a(x,\q,\bar\q) = \rme^{\d_\q}{\l}_\a \ .
\end{align}
Using (\ref{AV:SUSY}), 
the actions of the supercovariant derivatives on this superfield are \cite{WB}
\begin{align} \label{AV_Constraints}
	D_\a\L_\b = \frac1\k\eps_{\b\a} + \rmi\k \bar\L_\da\pd_\a^\da\L_\b\,,
	\qquad 
	\Db_\da\L_\b = -\rmi\k\L^\a\pd_\ada\L_\b \ .
\end{align}
The projection to the components of \eqref{SW-solns2} immediately reproduces 
the results of \cite{Rocek1978} and gives the relation between 
the constrained superfield Goldstino models and the (chiral) AV Goldstino.

${}$For the complex linear superfield $\S$,
the solution to the constraints \eqref{mCL:constraint}, \eqref{1st constraint} 
and \eqref{2nd constraint} in terms of $\bar \X_\da$ 
is very simple:
\begin{align} \label{mCL-SW-soln}
	2f\S = \bar{\X}^2 \ .
\end{align}
Projection to components yields $\r_\a=\x_\a$ and the component solutions
\eqref{solns-to-both-constraints}. So we see that the model proposed in this
paper is the natural constrained superfield to associate with the chiral AV 
Goldstino and the Samuel-Wess superfield \eqref{chiral-Goldstino-SF} 
can be considered derivative \eqref{SW_as_Derivative}.
The Ro\v{c}ek and Lindstr\"om-Ro\v{c}ek superfields can both be constructed
from the complex linear scalar as
\begin{align}
	\F = -\frac12 f\k^2 \Db^2(\bar\S \S) \quad\text{and}\quad
	V  = 2 f \k^2 \bar\S \S \ .
\end{align}

Unlike the chiral and real superfield cases, the solution of the
complex linear constraints in terms of the superfield $\L_\a$ is
different from that using $\X_\a$. Some work gives
\begin{align} \label{mCL-AV-soln}
\begin{aligned}
	2f\S &= 4\big(\bar\L^2+\frac\k2 D^\a(\L_\a\bar\L^2) 
			- \frac{\k^2}{16} D^2(\L^2\bar\L^2)\big)  \\
		 &= \bar\L^2\big(1-\rmi\k^2(\L\s^a\pd_a\bar\L)
		 	+ \k^4\L^2(\pd_a\bar\L\st^{ab}\pd_b\bar\L)\big) \,.
\end{aligned}
\end{align}	

\section{Couplings to matter and supergravity}\label{sect:couplings}
Complex linear superfields are ubiquitous in $\cN=2$ supersymmetry
in the sense that any off-shell $\cN=2$ hypermultiplet 
without intrinsic central charge contains a complex linear scalar 
as one of its $\cN=1$ components, see e.g.\ \cite{K-Srni} for a review.
This is one of the reasons to believe that the construction 
presented in this paper is of interest. 

The constraints \eqref{mCL:constraint} 
and \eqref{2nd constraint} admit nontrivial generalizations such as 
\begin{align}
\label{coupled mCL 1}
 	-\frac14 \Db^2 \S &= X\,, \qquad {\bar D}_\da X =0\,,  \\
\label{coupled mCL 2}
	-\frac14 \S\Db^2D_\a\S &= X D_\a\S\,, 
\end{align}
for some (composite) chiral scalar $X$
possessing a non-vanishing expectation value. 
Such constraints%
\footnote{Modified linear constraints of the form 
(\ref{coupled mCL 1}) were first introduced in \cite{DG} 
and naturally appear, e.g., when one considers ``massive'' 
off-shell $\cN=2$ sigma-models \cite{Kuzenko:2006nw} 
in projective superspace \cite{LR-projective}.
}
are compatible with the nilpotency condition \eqref{1st constraint}.  
This makes it possible to construct couplings of the Goldstino to matter fields. 
For example, 
we can choose $X = f + G_1(\vf) + G_2(\vf) \tr(W^\a W_\a)$, 
where $G_1$ and $G_2$ are arbitrary holomorphic functions of 
some matter chiral superfields $\vf$,
$W_\a$ is the field strength of a vector multiplet 
and the trace is over the gauge indices. 
The resulting Goldstino-matter couplings can be compared 
with those advocated recently by Komargodski and Seiberg \cite{KS}. 
In the approach of \cite{KS}, the Goldstino is described by 
a chiral superfield $\F$ subject to the nilpotent constraint 
(\ref{Roceks_Constraint1}).
Matter couplings for the Goldstino in \cite{KS}  are generated simply 
by adding suitable interactions to the Lagrangian.\footnote%
{The complex auxiliary field $F$ contained in $\F$ 
is to be eliminated using its resulting equation of motion, 
which renders the supersymmetry {\it on-shell}.}
In our case, the Goldstino superfield $\S$ also obeys 
the nilpotency condition $\S^2=0$, along with the differential constraints 
\eqref{mCL:constraint} and \eqref{2nd constraint}.
Matter couplings can be generated by deforming the latter constraints to 
the form given by eqs.\ \eqref{coupled mCL 1} and \eqref{coupled mCL 2}.
Similarly to the analysis in section \ref{sect:Constraints}, 
the constraints \eqref{1st constraint} and (\ref{coupled mCL 1})
can be solved in terms of the Goldstino ${\bar \rho}_\da$ 
and two more independent fields $U_{\a \da} $ and  ${\bar \chi}_\da$. 
The latter fields become functions of the Goldstino and matter fields 
upon imposing the constraint (\ref{coupled mCL 2}). 
The supersymmetry remains {\it off-shell}! 

We also note that the constraints 
\eqref{coupled mCL 1} and \eqref{coupled mCL 2}
can be further generalized to allow for 
a coupling to an Abelian vector multiplet;
this requires replacing the covariant derivatives in 
\eqref{coupled mCL 1} and \eqref{coupled mCL 2}
by gauge-covariant ones and turning $X$ into a covariantly chiral superfield,
with $X$ and $\S$ having the same U(1) charge.

The constraints \eqref{mCL:constraint} and \eqref{2nd constraint} 
can naturally be generalised to supergravity%
\footnote{Our conventions for $\cN=1$ supergravity correspond to \cite{BK}.}
as
\begin{align}
\label{mCL:constraint SuGra}
 	-\frac14 (\cDb^2 - 4R) \S &= X\,, \qquad 	\cDb_\da X = 0\,,  \\
\label{2nd constraint SuGra}
	-\frac14 \S  (\cDb^2 - 4R) \cD_\a\S &= X \cD_\a\S\ , 
\end{align}
for some covariantly chiral scalar $X$. 
Here $\cD_A =(\cD_a , \cD_\a , \cDb^\da)$ 
denote the superspace covariant derivative corresponding to 
the old minimal formulation \cite{old} for $\cN=1$ supergravity, 
and $R$ the covariantly chiral scalar component of the superspace torsion  
described in terms of $R$, $G_{\a \da}$ and $W_{\a \b \g}$
(see \cite{BK,GGRS,WB} for reviews). 
The constraints \eqref{mCL:constraint SuGra} and \eqref{2nd constraint SuGra} 
have to be accompanied by the nilpotency condition \eqref{1st constraint}. 
As an example, consider the simplest case when $X$ is constant. 
We represent $X = (\sqrt2 \k)^{-1} = \const$, 
where $\k$ can be chosen to be real.
As a minimal generalization of (\ref{SW_as_Derivative}), 
we now introduce spinor superfields
$\X_\a = \frac{1}{\sqrt2}\cD_\a \bar \S$ and 
$	\bar\X_\da = \frac{1}{\sqrt2} \bar\cD_\da\S $.
Using the constraints (\ref{1st constraint}), (\ref{mCL:constraint SuGra}) 
and (\ref{2nd constraint SuGra}), 
we can derive closed-form constraints obeyed, e.g., by ${\bar\X}_\da$. 
They are
\begin{align}
\label{SW SuGra 1}
	\cDb_\da {\bar \X}_\db 	&=  \eps_{\da\db} 
		\Big( \frac1\k -\k R \,{\bar \X}^2 \Big) \, , \\
\label{SW SuGra 2}
	\cD_\a {\bar \X}_\da &=  \k \Big( 2\rmi  \,{\bar \X}^\db \cD_{\a \db} 
		{\bar \X}_\da  -G_{\a \da } {\bar \X}^2 \Big)\ ,
\end{align}
where $G_{\a\da}$ is the supergravity extension of the traceless Ricci tensor 
(see \cite{BK,GGRS,WB} for more details). 
The constraints (\ref{SW SuGra 1}) and (\ref{SW SuGra 2}) were introduced by 
Samuel and Wess \cite{SamuelWess1983} as a result of the nontrivial guess work
(these constraints are non-minimal generalizations of 
(\ref{SW_Constraint1}) and (\ref{SW_Constraint2})). 
In our approach, these constraints are trivial consequences 
of the formulation in terms of the complex linear Goldstino superfield.

We believe that our results will provide a useful contribution 
to the existing literature on Goldstino couplings to 
supersymmetric matter and supergravity, see 
\cite{SamuelWess1983, Ivanov, KS, Samuel:1983jp, Clark:1996aw, Luo:2010zp} 
and references therein.
\\

\noindent
{\bf Acknowledgements:}\\
We are grateful to Ulf Lindstr\"om for comments on the manuscript.
The work of SMK is supported in part by the Australian Research Council.



\begin{footnotesize}
\providecommand{\href}[2]{#2}
\begingroup\raggedright
\endgroup
\end{footnotesize}

\end{document}